\documentclass[%
 reprint,
 amsmath,amssymb,
 aps,
]{revtex4-1}

\usepackage{xcolor}
\usepackage{graphicx}
\usepackage{dcolumn}
\usepackage{bm}
\usepackage{float} 

\usepackage[utf8]{inputenc}
\usepackage[english]{babel}
\usepackage{graphicx}
 \graphicspath{{figures/}{../figures/}}
\newcommand*{\affaddr}[1]{#1} 
\newcommand*{\affmark}[1][*]{\textsuperscript{#1}}

\newcommand{\beq}{\begin{equation}}
\newcommand{\eeq}{\end{equation}}
\newcommand{\be}{\begin{equation}}
\newcommand{\ee}{\end{equation}}
\newcommand{\bea}{\begin{eqnarray}}
\newcommand{\eea}{\end{eqnarray}}

\usepackage{subfiles}
 
\usepackage{blindtext}
 
\begin{document}
 
 \preprint{APS/123-QED}

\title{Anapole Dark Matter via Vector Boson Fusion Processes at the LHC} 

\author{
Andr\'es Fl\'orez\affmark[2], Alfredo Gurrola\affmark[1], Will Johns\affmark[1], Jessica Maruri\affmark[1], Paul Sheldon\affmark[1], Kuver Sinha\affmark[3], Savanna Rae Starko\affmark[1]\\
\affaddr{\affmark[1] Department of Physics and Astronomy, Vanderbilt University, Nashville, TN, 37235, USA}\\
\affaddr{\affmark[2] Physics Department, Universidad de los Andes, Bogot\'a, Colombia}\\
\affaddr{\affmark[3] Department of Physics and Astronomy, University of Oklahoma, Norman, OK, 73019, USA}
}

\date{\today}
             
\begin{abstract}

Dark matter that is electrically neutral but couples to the electromagnetic current through higher-dimensional operators constitutes an interesting class of models. We investigate this class of models at the Large Hadron Collider, focusing on the anapole moment operator in an effective field theory (EFT) framework, and utilizing the vector boson fusion (VBF) topology. Assuming proton-proton collisions at $\sqrt{s} = 13$ TeV, we present the VBF anapole dark matter (ADM) cross sections and kinematic distributions as functions of the free parameters of the EFT, the cutoff scale $\Lambda$ and the ADM mass $m_{\chi}$. 
We find that the distinctive VBF topology of two forward jets and large dijet pseudorapidity gap 
is effective at reducing SM backgrounds, leading to a $5\sigma $ discovery reach for all kinematically allowed ADM masses with 
$\Lambda \le 1.65$ (1.15) TeV, assuming an integrated luminosity of 3000 (100) fb$^{-1}$. 

\end{abstract}


\maketitle


\section{\label{sec:level1}Introduction}

Determining the identity of dark matter (DM) is one of the most active areas of research in particle physics. One interesting class of DM models is where the DM particle is itself electrically neutral, but couples to the photon through higher multipole interactions. This scenario has been considered by many authors (\cite{Raby:1987ga} - \cite{Barger:2010gv}) in a variety of UV settings: technicolor \cite{Bagnasco:1993st, Banks:2010eh}, composite dark sectors \cite{Antipin:2015xia}, supersymmetry \cite{Dutta:2014jda}, simplified leptophilic models \cite{Kopp:2014tsa, Sandick:2016zut}, and simplified light dark sectors \cite{Chu:2018qrm}. Different multipoles have been studied, including electric and magnetic dipole moments, the anapole moment, and charge radius interaction.

The purpose of this paper is to probe multipole moments at the Large Hadron Collider (LHC) using the vector boson fusion (VBF) topology (\cite{VBFHiggsTauTauCMS} - \cite{VBFHN}). We focus on the anapole moment, leaving other moments for future work. We work within an effective field theory framework, remaining agnostic about the UV completion. The anapole dark matter (ADM) operator can be written as
\begin{equation} 
\label{anapoleEFT}
\mathcal{L}_\text{eff,anapole} = \frac{g}{\Lambda^2} \, \bar\chi \gamma^\mu \gamma^5 \chi \, \partial^\nu F_{\mu\nu} \,\,,
\end{equation}
where $\Lambda$ is the cutoff scale and $\chi$ denotes the DM particle. Possible UV completions could be Bino DM coupling to sleptons in supersymmetry or DM that is a composite state of charged particles (where $\Lambda$ would be the confinement scale).

Collider studies of this class of operators have typically relied on the mono-$X$ signature, where $X$ can be a jet \cite{Gao:2013vfa}, a $Z$-boson~\cite{Alves:2017uls} - \cite{Carpenter:2012rg}, or a photon \cite{Primulando:2015lfa}. In a recent paper \cite{Alves:2017uls}, one of the authors studied the effective operator in Eq.~\ref{anapoleEFT} using the mono-$Z$ signature  at the high-luminosity LHC (HL-LHC) \cite{Alves:2015dya, Carpenter:2012rg}. The discovery potential of the HL-LHC was determined using boosted decision trees for various levels of systematic uncertainties. The authors of \cite{Primulando:2015lfa} studied magnetic dipole moment operators using monojet, monophoton, and diphoton searches at the LHC, 100 TeV collider, and the ILC. In both these papers, a comparison to the projected bounds from direct detection experiments was performed. We refer to \cite{Gao:2013vfa}  and references therein for some older studies.

VBF provides a strategy in this context that is complementary to the above searches. As we discuss later, the cross section of VBF ADM dominates over the cross section of mono-$Z$ for all relevant values of $m_\chi$ and $\Lambda$. Moreover, while a mono-$Z$ study in this context has to contend with irreducible Standard Model (SM) $ZZ$ and $W^+W^-$ backgrounds, the VBF topology offers remarkable control over SM backgrounds. This control is due to the presence of two distinctive forward energetic jets, in opposite hemispheres, with large dijet invariant mass. A comparative study of VBF with other mono-$X$ searches for multipole DM would be interesting in the future.

We note that several of the current authors have exploited these attractive features of VBF processes to propose effective LHC probes of WIMP DM in models of Supersymmetry (SUSY)~\cite{DMmodels2}, SUSY electroweakinos~\cite{VBF1}, SUSY sleptons~\cite{VBFSlepton}, SUSY top and bottom squarks in compressed spectra
~\cite{VBFStop,VBFSbottom}, $Z^{'}$~\cite{VBFZprime}, new heavy spin-2 bosons~\cite{VBFY2}, and heavy neutrinos~\cite{VBFHN}. Although triggering, reconstructing, identifying, and calibrating a pair of forward jets presents an experimental challenge, some of the proposed searches for SUSY WIMP DM and compressed spectra have been successfully carried out by the CMS collaboration~\cite{CMSVBFDM,VBF2}.

The rest of our paper is structured as follows. In Section \ref{sample}, we describe our simulation methods, signal cross section, and dominant backgrounds. In Section \ref{criteria}, we discuss event selection criteria, and in Section \ref{results}, main results. We end with a short discussion in Section \ref{discussion}.

\section{Samples and simulation} \label{sample}

Simulated events from proton-proton ($pp$) collisions at $\sqrt{s}=13$ TeV were generated for signal and background using MadGraph5\_aMC (v2.6.3)~\cite{MADGRAPH}. Hadronization was performed with PYTHIA (v6.416) \cite{Sjostrand:2006za}. Detector effects were included through Delphes (v3.3.2) \cite{deFavereau:2013fsa}, using the CMS input card. The signal model was produced using FeynRules~\cite{FeynRules}, following Ref.~\cite{Alves:2017uls}. We produced several signal samples considering various values of $\mathcal{A}\equiv \frac{g}{\Lambda^2}$, which as expected has a direct impact on the  production cross section.

\begin{figure}
\begin{tabular}{c}
    \includegraphics[scale = 0.5]{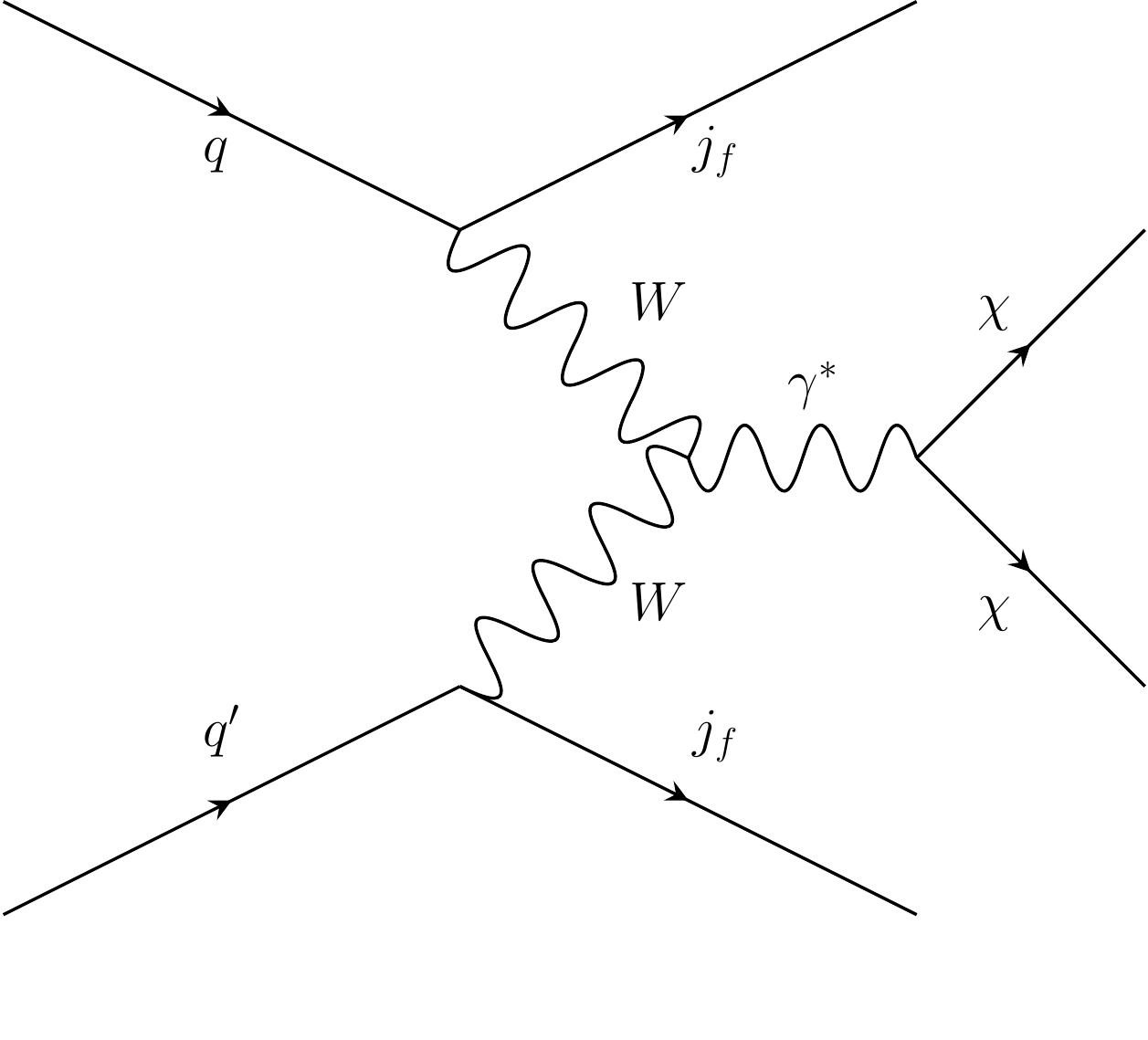}
    \\
    \includegraphics[width = .35\textwidth, height = .25\textheight]{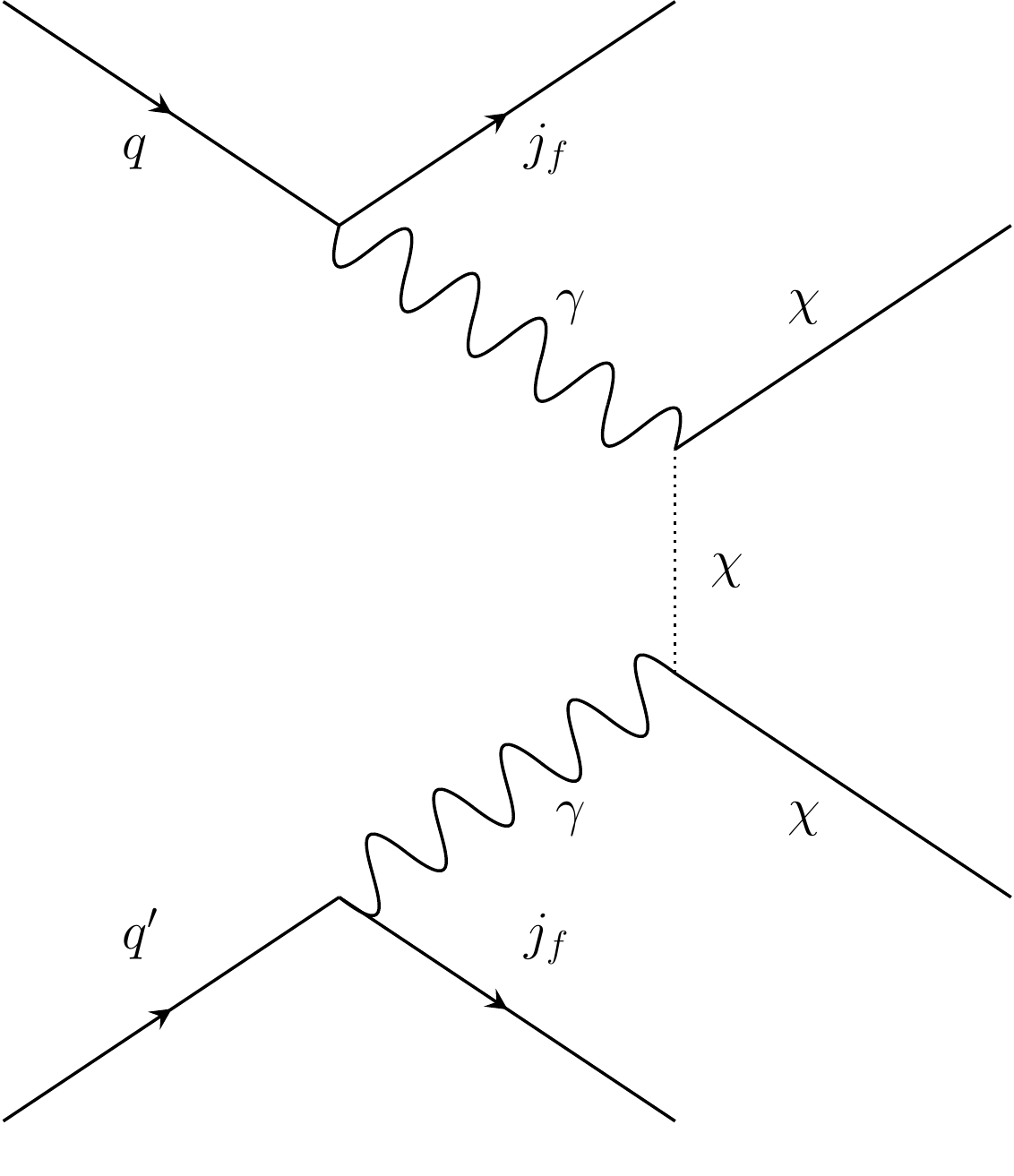}
\end{tabular}
    \caption{Representative Feynman diagrams depicting VBF $\chi\chi jj$ production.}
    \label{fig:feyn}
\end{figure}

\begin{figure}
    \centering
    \includegraphics[width=.5\textwidth, height=.25\textheight]{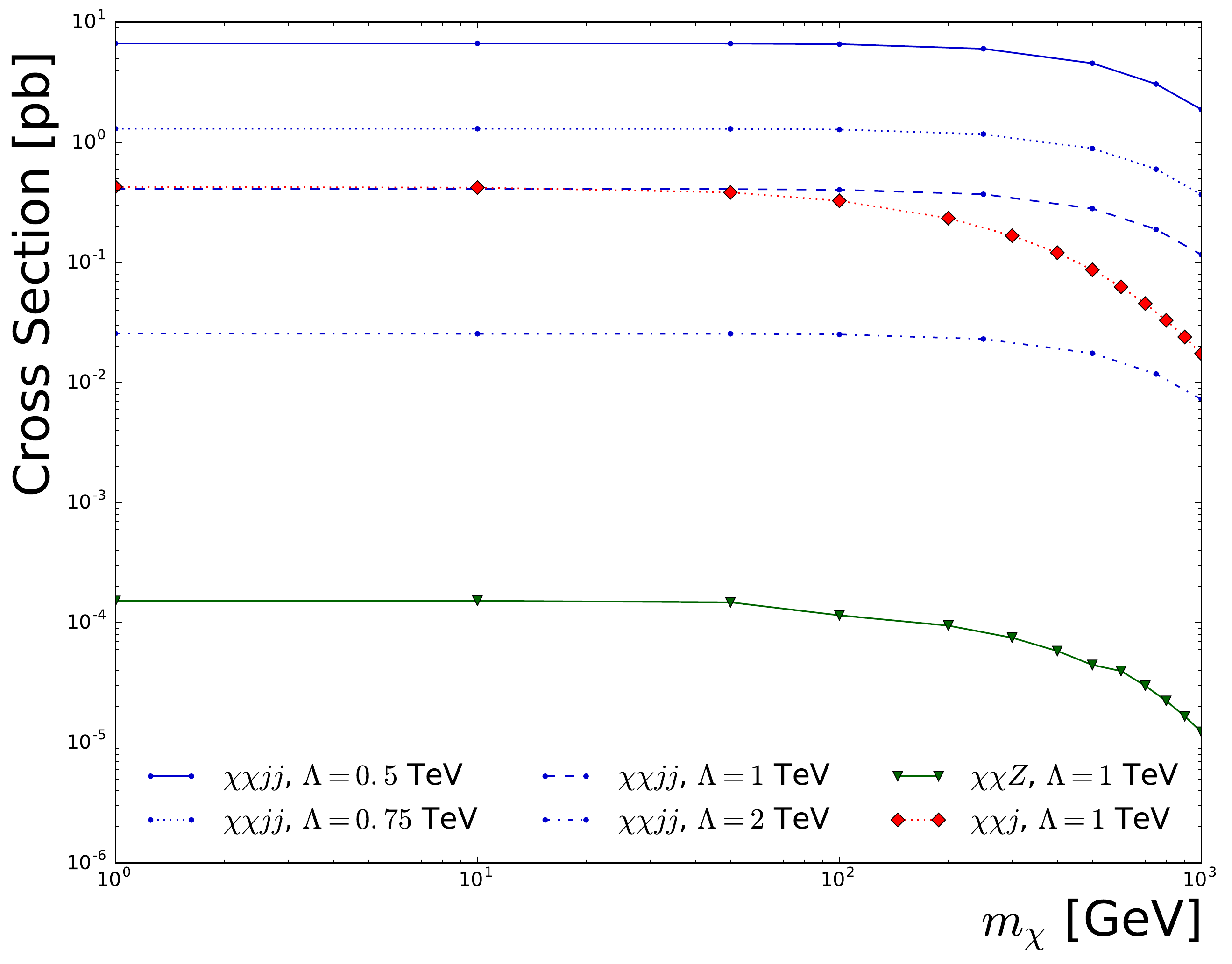}
    \caption{The VBF $\chi\chi jj$ cross section as a function of $\Lambda$ and $m_{\chi}$. For comparison, the mono-jet and mono-$Z$ cross sections are also shown.}
    \label{VBFcrosssection}
\end{figure}

\indent Signal samples were produced for a variety of ADM masses, ranging from 1 GeV to 3000 GeV (1, 10, 50, 100, 250, 500, 750, 1000, 2000, and 3000 GeV). The value of $\Lambda$ was varied between 500 GeV to 3000 GeV, in steps of 10 GeV, for every ADM mass point considered. The signal samples were produced considering pure electroweak production of a $\chi\chi$ pair and two additional jets (i.e. $pp\rightarrow \chi\chi jj$ with suppressed QCD coupling $\alpha_{QCD}^{0}$). Figure~\ref{fig:feyn} shows representative Feynman diagrams depicting VBF $\chi\chi jj$ production.  Figure~\ref{VBFcrosssection} shows the VBF ADM production cross section as a function of $m(\chi)$ for varying values of $\Lambda$. As expected, the cross section scales as $\Lambda^{-4}$. Interestingly, we find that the VBF ADM production cross sections dominate over those of the more traditional mono-$Z$ and monojet processes for all relevant values of $\Lambda$ and $m_{\chi}$, making VBF an important mode for discovery.

The dominant sources of SM background are production of a $Z$ or $W$ boson with associated jets, referred to as $Z$+jets and $W$+jets, respectively. The  $Z$+jets and $W$+jets backgrounds together constitute about 95\% of the total SM background. Due to the genuine missing momentum from neutrinos, $W(\to\ell\nu)$+jets events become an important background if the accompanying charged lepton is ``lost'' either because it falls outside the geometric acceptance of the detector or fails the lepton identification criteria (and thus fails the lepton veto criteria described later). The $Z$+jets process becomes an important and irreducible background when the missing momentum arises from $Z$ boson decays to neutrinos. Finally, around a 5\% contribution from $t\bar{t}+$jets events is also expected. Similar to the $W$+jets process, $t\bar{t}+$jets becomes a background when leptons from the $t\to W\to\ell\nu$ decays are ``lost'' and the bottom quarks fail the b-jet identification criteria. These major SM backgrounds were produced considering up to four additional jets associated to the central process, inclusive in the electroweak coupling ($\alpha_{EWK}$) and $\alpha_{QCD}$. At parton level the jets were required to have a transverse momentum ($p_{\textrm{T}}$) above 20 GeV and pseudorapidity ($|{\eta}|$) $|{\eta}| < 5$.

Jet matching was included using the MLM algorithm \cite{MLM}. The matching requires an optimization of the xqcut and qcut variables in the algorithm, which for this letter are set to 15 and 35 GeV respectively. The xqcut is defined as the minimal distance among partons at generation level, and the qcut corresponds to minimum energy spread for a clustered jet in PYTHIA. The optimization was performed using the differential jet rate distribution included in MadGraph. The distribution includes events with different jet multiplicities and the optimal parameters must result in a smooth transition between the corresponding curves for events with $n-1$ and $n$ jets. 

\section{Event selection criteria} \label{criteria}
\begin{figure}
    \centering
    \includegraphics[width=0.45\textwidth, height=0.25\textheight]{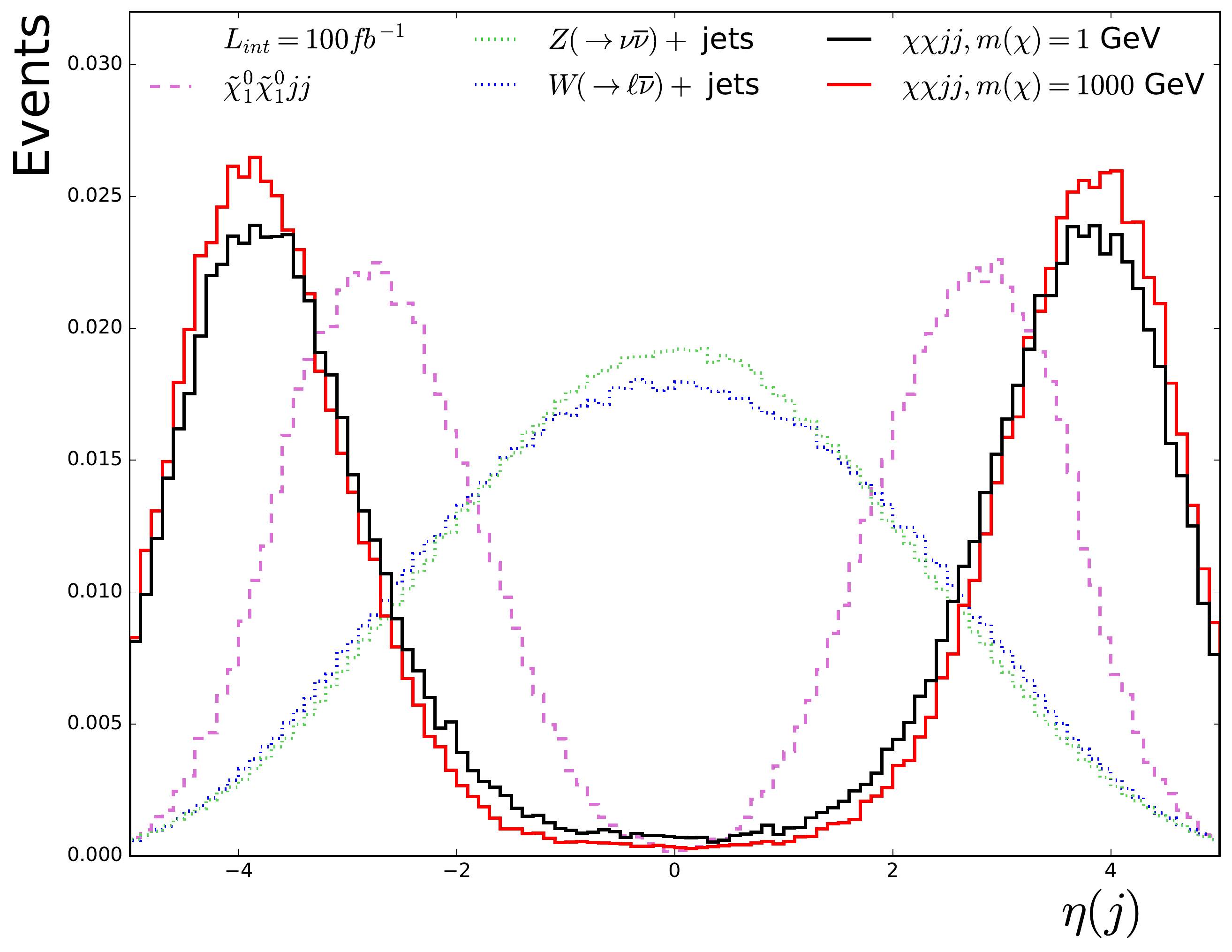}
    \caption{$\eta(j)$ distributions (normalized to unity) for the major SM backgrounds (blue and green), VBF neutralino $\tilde{\chi}_{1}^{0}$ pair production in SUSY (purple), and the benchmark signal samples with $\{\Lambda, m_{\chi}\} = \{1000$ GeV, $1$ GeV$\}$ (black) and $\{1000$ GeV, $1000$ GeV$\}$ (red).}
    \label{eta_distribution}
\end{figure}

\begin{figure}
    \centering
    \includegraphics[width=0.45\textwidth,height=0.25\textheight]{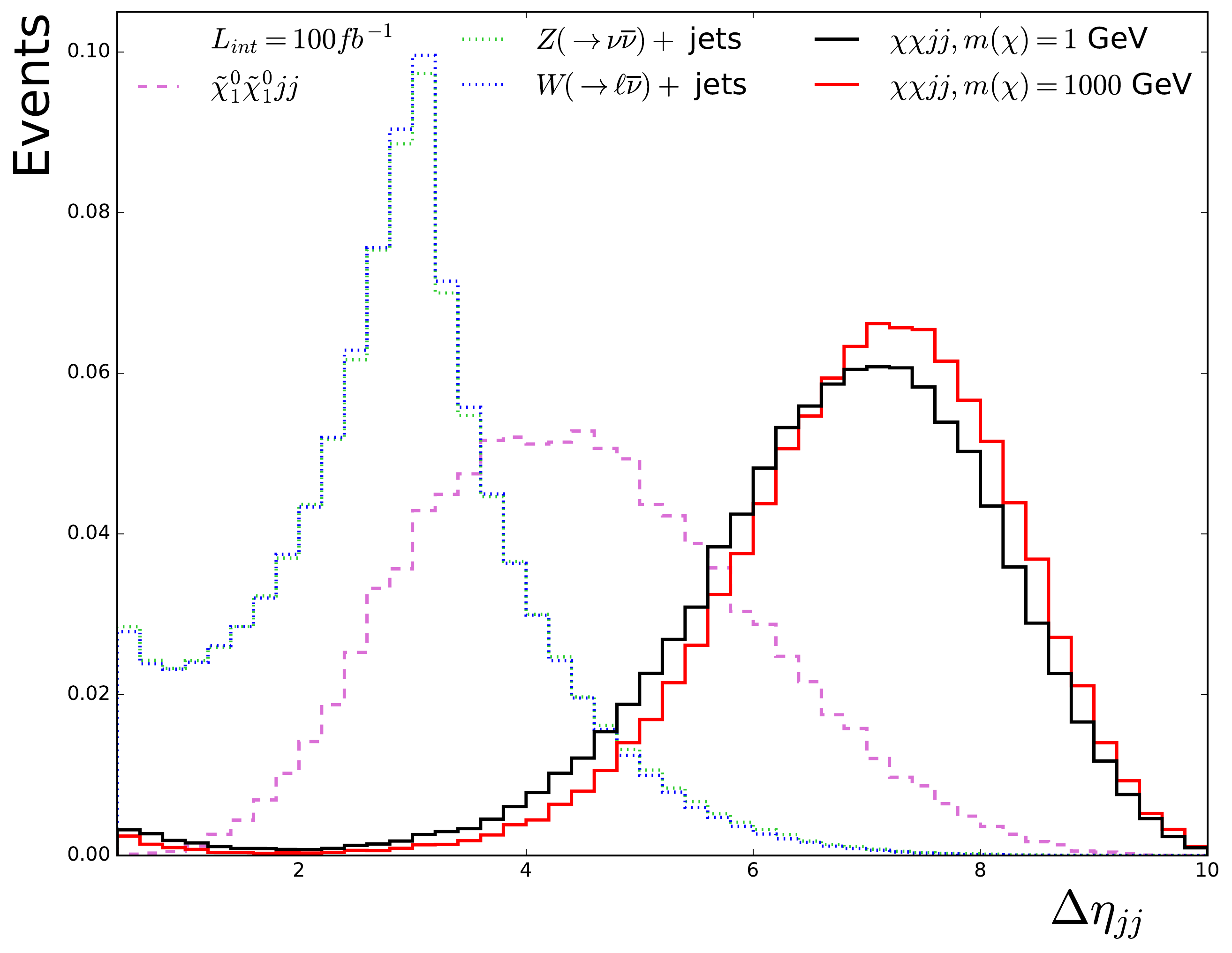}
    \caption{$\Delta\eta_{jj}$ distributions (normalized to unity) for the major SM backgrounds (blue and green), VBF neutralino $\tilde{\chi}_{1}^{0}$ pair production in SUSY (purple), and the benchmark signal samples with $\{\Lambda, m_{\chi}\} = \{1000$ GeV, $1$ GeV$\}$ (black) and $\{1000$ GeV, $1000$ GeV$\}$ (red).}
    \label{delta_eta_distribution}
\end{figure}

The VBF topology is characterized by a pair of high $p_{T}$ forward jets located in opposite hemispheres of the detector.  Since the minimum $p_{T}$ of reconstructed jets is limited by experimental constraints, namely from the detector geometry, detector performance, and jet reconstruction algorithms, we select events with at least two jets with $p_{T}(j)>30$ GeV.  Figure~\ref{eta_distribution} shows the $\eta$ distribution of jets for our major SM backgrounds and two signal benchmark samples with $\Lambda = 1$ TeV and $m_{\chi} = 1$ GeV (1 TeV) in black (red).  Similarly, Figure~\ref{delta_eta_distribution} displays a comparison of the pseudorapidity difference $|\Delta\eta_{jj}|$ between the two leading jets in the signal and background samples.  While the SM background contributions typically consist of events containing central jets ($\eta\approx 0$) and dijet pairs with small $|\Delta\eta_{jj}|$, the ADM signature is characterized by jets with $\eta\approx 4$ and large $|\Delta\eta_{jj}|$.  Therefore, Figures~\ref{eta_distribution} and~\ref{delta_eta_distribution} motivate a forward $\eta$ requirement on the two leading jets and a large $\eta$ gap between them to effectively differentiate signal from background.

We note that VBF ADM production is fundamentally different from the VBF SUSY processes studied by some of the current authors in Refs.~(\cite{DMmodels2} - \cite{VBFSbottom}). While VBF SUSY production occurs via t-channel $WW$/$WZ$/$ZZ$ diagrams, an important VBF production mechanism in the specific case of ADM is t-channel $\gamma\gamma$ fusion (see Figure~\ref{fig:feyn}).  This distinguishing feature of VBF ADM production results in significantly more forward jets and a larger $\Delta\eta_{jj}$ gap than those found in events from other VBF processes and DM scenarios.  This difference allows for the experimental differentiation of this ADM process from other DM scenarios.  In addition to the SM background and VBF ADM $\eta(j)$ and $|\Delta\eta_{jj}|$ distributions, Figures \ref{eta_distribution} and \ref{delta_eta_distribution} also display the corresponding $\eta(j)$ and $|\Delta\eta_{jj}|$ distributions for VBF neutralino ($\tilde{\chi}_{1}^{0}$) pair production in SUSY in order to elucidate this point.

The event selection thresholds were determined using an optimization process for the best signal significance $z$.  The signal significance was determined based on the signal-to-noise ratio, which is given explicitly by $z = \frac{S}{\sqrt{S+B+(.25B)^{2}}}$ where $S$ is the signal event yield and $B$ is the combined background event yield. This signal significance calculation includes a 25$\%$ systematic uncertainty (described later), which is standard for VBF searches at CMS ~\cite{CMSVBFDM,VBF2}.  The samples used for selection optimization were $m_{\chi}$ = 1 GeV, $m_{\chi}$ = 500 GeV, and $m_{\chi}$ = 1000 GeV with $\Lambda = 1$ TeV in all $m_{\chi}$ considerations.  The selection thresholds were each chosen such that signal significance was at the maximum.

\begin{figure}
     \centering
     \includegraphics[width=0.45\textwidth, height=0.25\textheight]{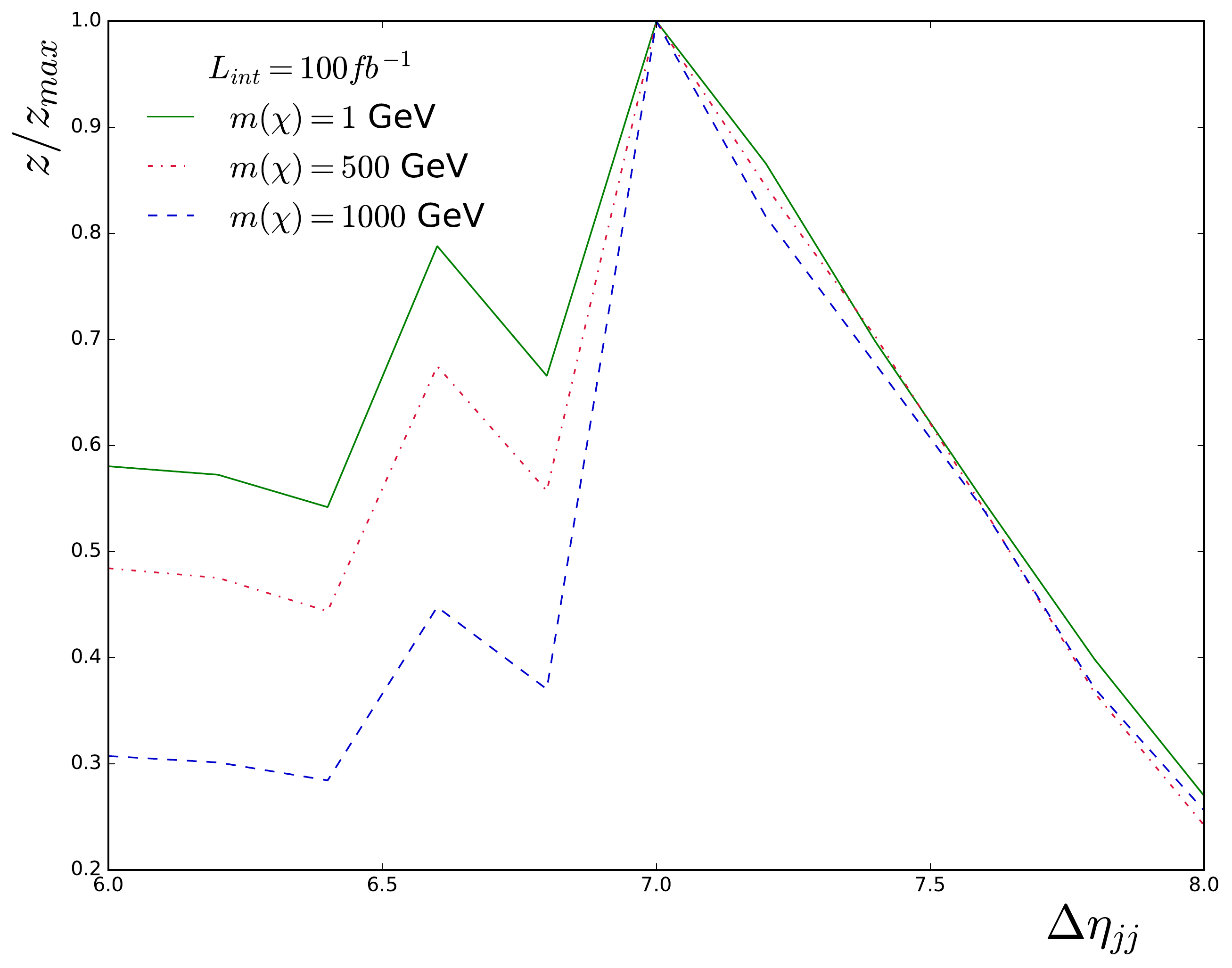}
     \caption{The normalized signal significance $z/z_{max}$ as a function of the $|\Delta\eta_{jj}|$ requirement.  The signal significance is optimized for $|\Delta\eta_{jj}| > 7.0$.}
     \label{deltaeta}
 \end{figure}

In Figure \ref{deltaeta}, we display the normalized signal significance $z/z_{max}$ as a function of $|\Delta\eta_{jj}|>X$ cut value, assuming an integrated luminosity of 100 $fb^{-1}$.  The signal significance is maximized for $|\Delta\eta_{jj}| > 7$.  Similarly, the optimization procedure leads us to require $|\eta(j)|>3.0$

 \begin{table}
     \centering
     \setlength{\tabcolsep}{1.5em}
     \def\arraystretch{1.25}
     \begin{tabular}{l l}
     \hline
    Criterion        & Selection    \\
    \hline
    $|\eta(j)|$      & $> 3.0$      \\
    $p_{T}(j)$       & $> 30$ GeV   \\
    $N(j)$           & $\geq 2$     \\
    $p_{T}(\ell)$    & $> 10$ GeV   \\
    $|\eta(\ell)|$   & $< 2.5$      \\
    $N(\ell)$        & $=0$         \\
    $p_{T}$(b-jet)   & $> 30$ GeV   \\
    $|\eta|$(b-jet)  & $< 2.4$      \\
    $N$(b-jet)       & $=0$         \\
    $\Delta\eta_{jj}$ & $ > 7.0$     \\
    $E_{T}^{miss}$   & $> 175$ GeV  \\
    \hline
     \end{tabular}
     \caption{Event selection criteria for the proposed VBF ADM search region.}
     \label{fullcuts}
 \end{table}

\begin{figure}
    \centering
    \includegraphics[width=0.45\textwidth, height=0.25\textheight]{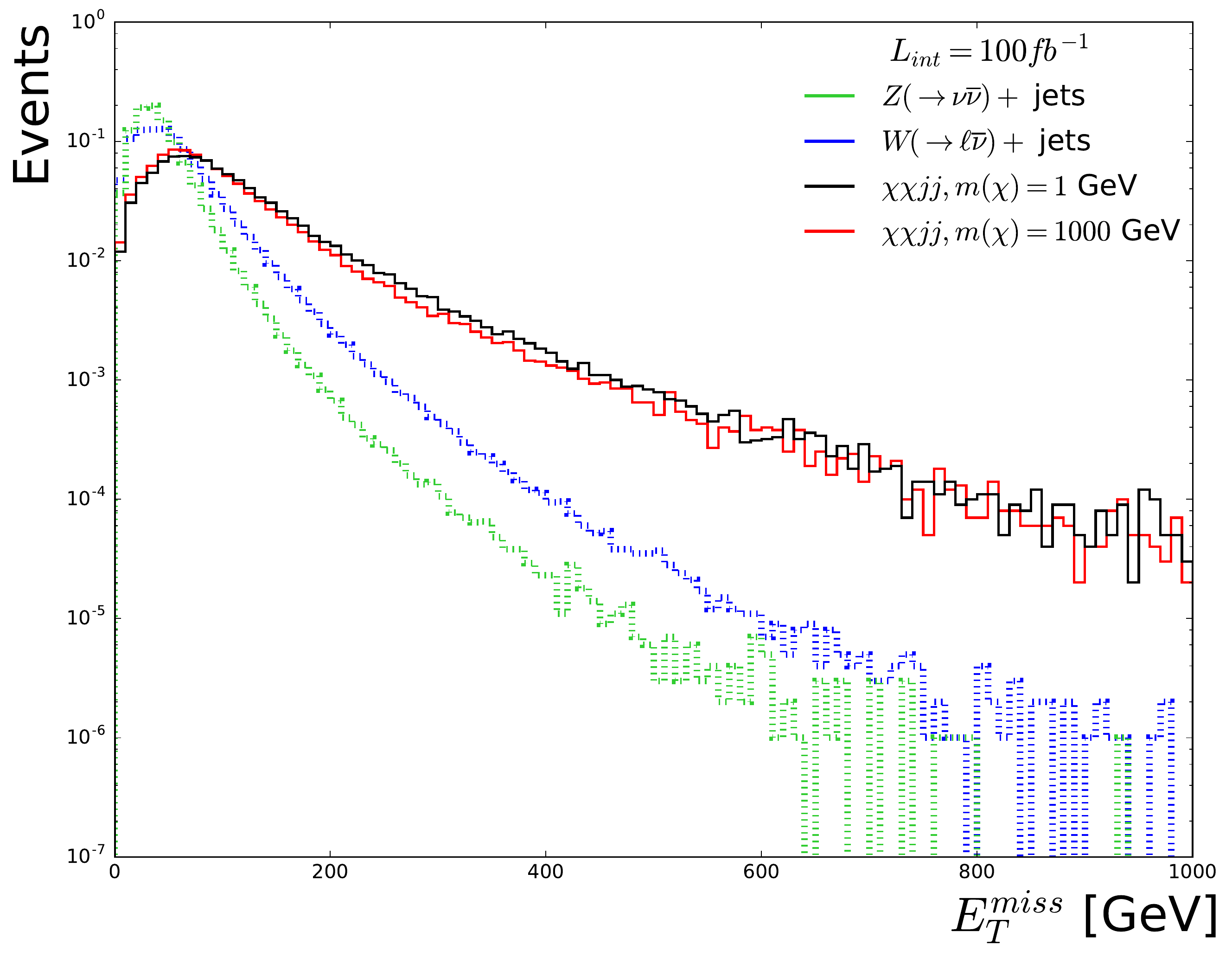}
    \caption{$E_{T}^{miss}$ distributions (normalized to unity) for the major SM backgrounds (blue and green) and the benchmark signal samples with $\{\Lambda, m_{\chi}\} = \{1000$ GeV, $1$ GeV$\}$ (black) and $\{1000$ GeV, $1000$ GeV$\}$ (red).}
    \label{met_distribution}
\end{figure}

\begin{figure} 
     \centering
     \includegraphics[width=0.45\textwidth, height=0.25\textheight]{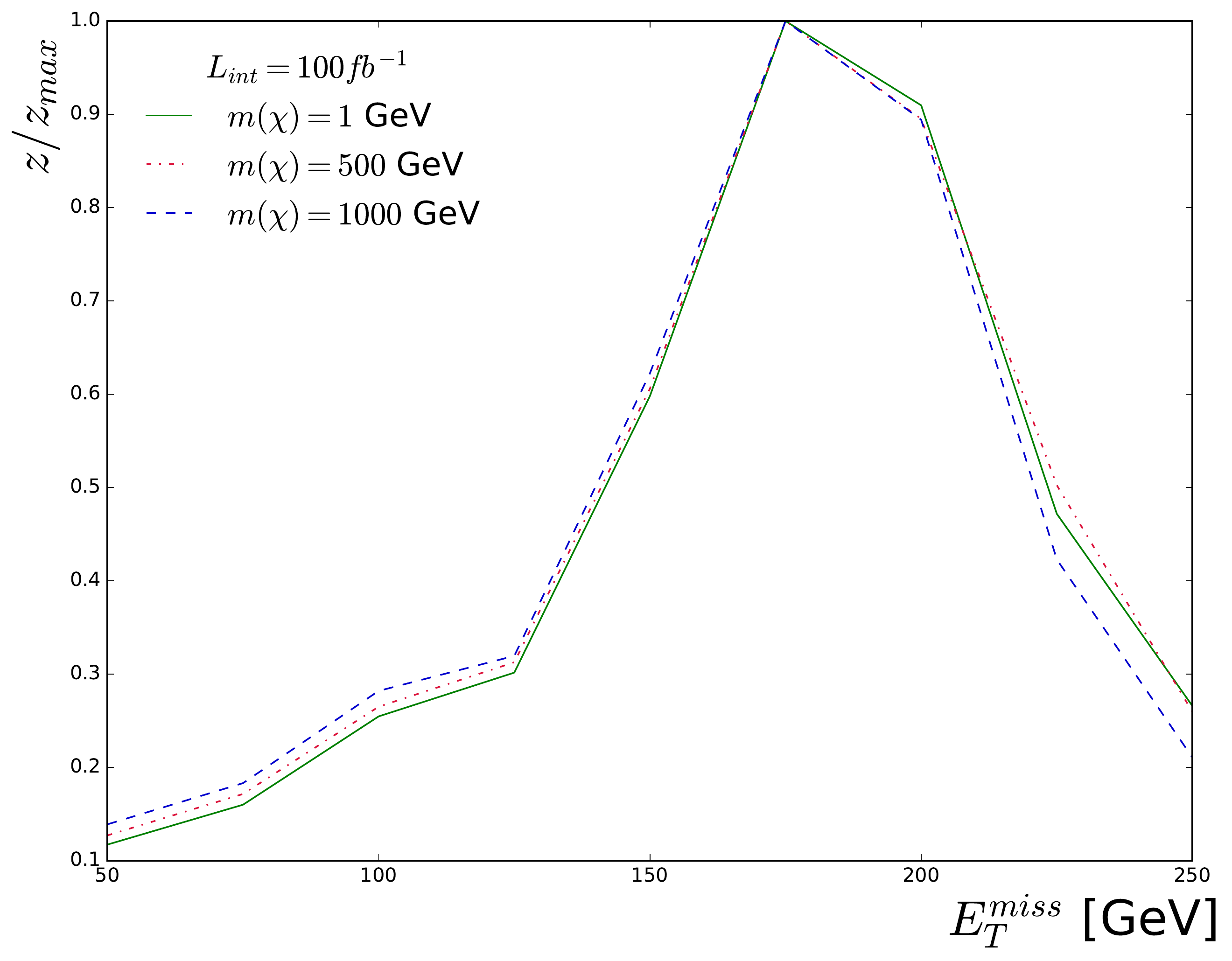}
     \caption{The normalized signal significance $z/z_{max}$ as a function of the $E_{T}^{miss}$ requirement.  The signal significance is optimized for $E_{T}^{miss} > 175$ GeV.}
     \label{met}
 \end{figure}

Similar to current ATLAS and CMS DM searches utilizing the mono-$X$ signature, the production of DM candidates at the LHC is indirectly inferred by measuring the imbalance of the total energy in the transverse plane of the detectors ($E_{T}^{miss}$).  The reconstructed ${E}_{T}^{miss}$ is the magnitude of the negative vector sum of the transverse momentum  of visible objects, $E_{T}^{miss} = |-\sum_{i=visible}\vec{p}_{T,i}|$, where $\vec{p}_{T,i}$ is the transverse momentum vector of all visible particles $i$ in an event.  Due to the neutrinos from $W/Z$ decays, the major SM backgrounds include genuine $E_{T}^{miss}$, which is constrained by the $Z$ or $W$ mass.  On the other hand, the ADM signature of interest has, on average, a harder $E_{T}^{miss}$ distribution that can be used to discriminate against the SM backgrounds.  Figure \ref{met_distribution} displays the $E_{T}^{miss}$ distributions (normalized to unity) for the major SM backgrounds and the benchmark signal samples.  Figure \ref{met} shows the normalized signal significance $z/z_{max}$ as a function of the $E_{T}^{miss}$ cut value, assuming an integrated luminosity of 100 $fb^{-1}$.  As previously mentioned, we choose the cut value that achieves maximum signal significance.  From Figure \ref{met}, it is evident that signal significance is maximized for $E_{T}^{miss} > 175$ GeV.

To further suppress SM backgrounds and to isolate the distinct VBF ADM signal, we impose $b$-jet and lepton veto selections.  Events are rejected if a jet has $p_{T} > 30$ GeV, $|\eta| < 2.4$, and is identified as a bottom quark ($b$).  With the b-jet veto, SM backgrounds with top quarks are suppressed.  Events are also rejected if they contain an identified electron or muon candidate with $|\eta| < 2.5$ and $p_{T} > 10$ GeV.  Similarly, simulated events in the proposed search region are required to have zero jets with $p_{T} > 20$ GeV and $|\eta|<2.5$ tagged as hadronically decaying tau leptons ($\tau_{h}$).  The b-jet and lepton veto selections also reduce SM backgrounds with vector boson pairs (e.g. $WW\to\ell\nu\ell\nu$) and $Z/\gamma^{*}\to\ell\ell$ to negligible values, while being $>90\%$ efficient for VBF ADM signal events.  Table \ref{fullcuts} summarizes the final optimized signal selection criteria.

\begin{figure} 
     \centering
     \includegraphics[width=0.45\textwidth, height=0.25\textheight]{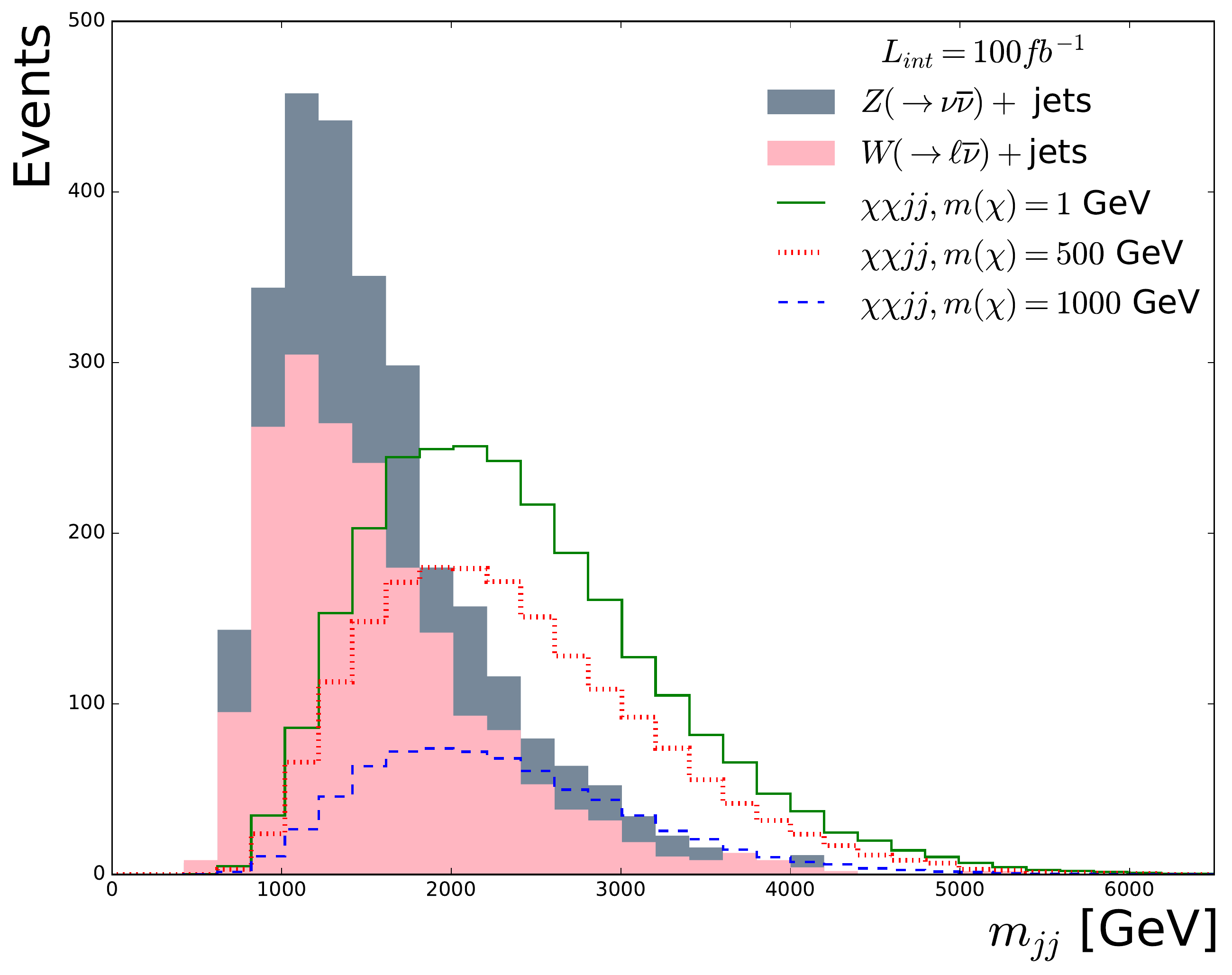}
     \caption{Dijet mass $m_{jj}$ distribution of VBF ADM signal benchmark samples and major SM backgrounds with all the optimized selections implemented, summarized in Table \ref{fullcuts}.}
     \label{dijetmass}
 \end{figure}

\begin{figure} 
     \centering
     \includegraphics[width=0.45\textwidth,height=0.25\textheight]{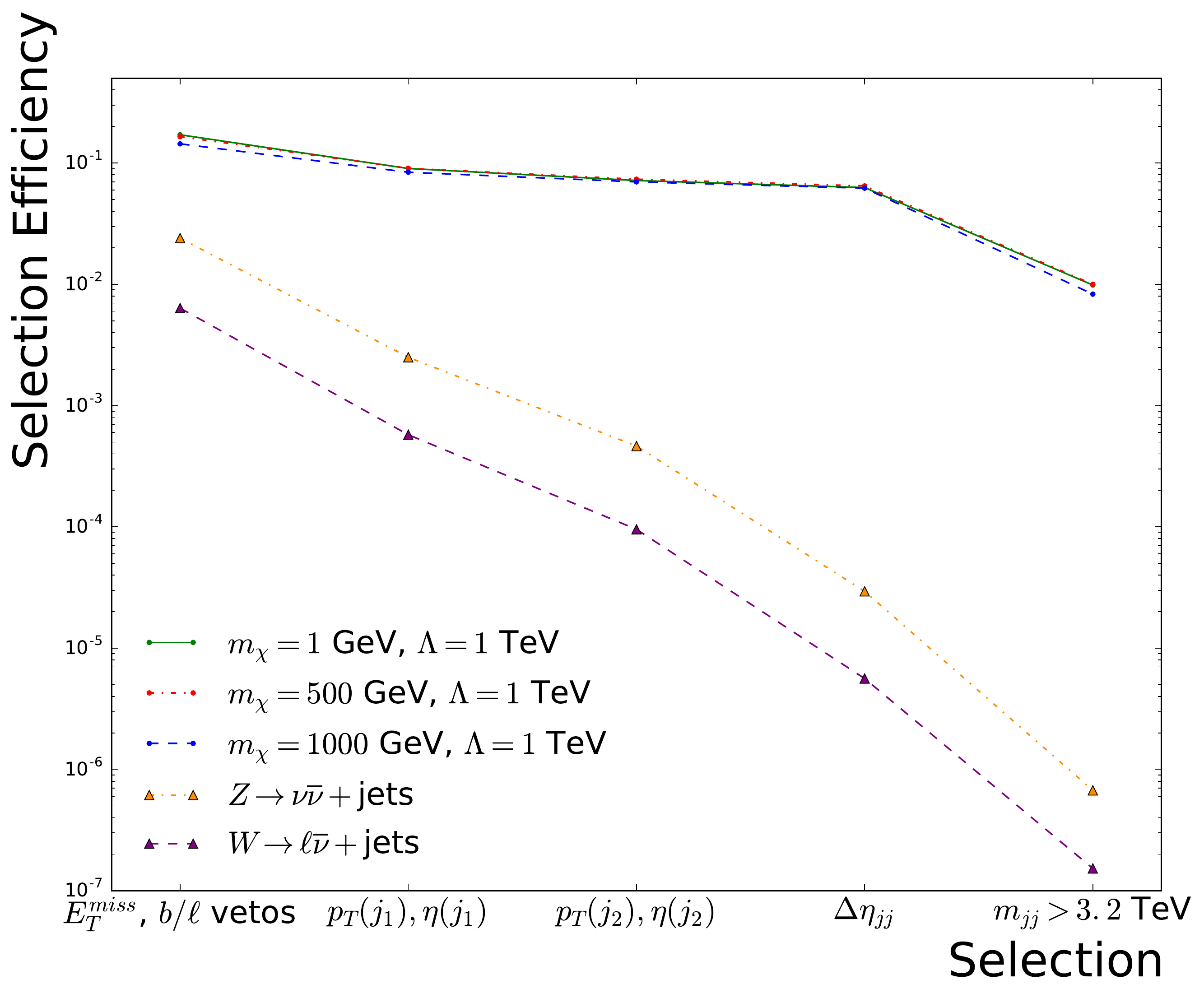}
     \caption{Selection efficiency after each additional criteria outlined in Table \ref{fullcuts}.  The benchmark signal signatures are the green, red, and blue curves while major SM backgrounds are the orange and purple curves.}
     \label{efficiency}
 \end{figure}

\begin{figure} 
     \centering
     \includegraphics[width=0.45\textwidth, height=0.25\textheight]{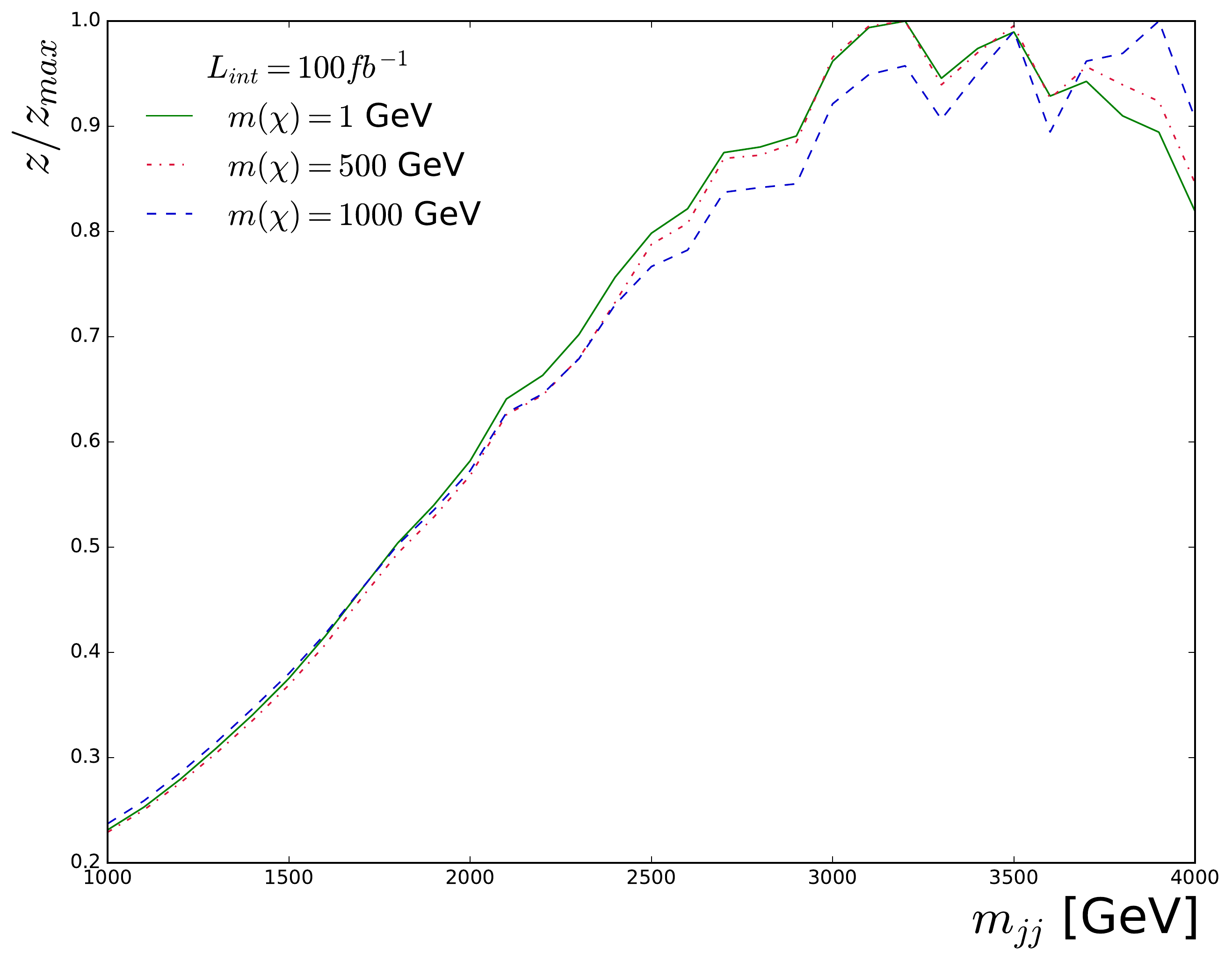}
     \caption{The normalized signal significance $z/z_{max}$ as a function of $m_{jj}$ cut value. Although final discovery potential is calculated using a shape based analysis of the full $m_{jj}$ range, this plot illustrates that the bulk of the sensitivity lies in the tail of the $m_{jj}$ spectrum (i.e. above 3 TeV).}
     \label{mjj_optimization}
 \end{figure}

The very forward jet requirement with unusually large $|\Delta\eta_{jj}|$ gap characterizing the VBF ADM topology results in TeV scale dijet mass $m_{jj}$.  The dijet mass is given by $m_{jj} = \sqrt{2p_{T}^{j_{1}}p_{T}^{j_{2}}cosh(\Delta\eta(jj))}$.  Figure \ref{dijetmass} shows the $m_{jj}$ distributions for the main SM backgrounds and the signal benchmark samples with $\{\Lambda, m_{\chi}\} = \{ 1000$ GeV, $1$ GeV$\}$, $\{ 1000$ GeV$, 500$ GeV$\}$, and $\{ 1000$ GeV$, 1000$ GeV$\}$. The distributions are obtained after applying the fully optimized selection criteria outlined in Table~\ref{fullcuts}, and normalized to the expected yields assuming an integrated luminosity of $L_{int} = 100$ fb$^{-1}$. The bulk of the background distribution lies at $m_{jj}$ values below 2 TeV, while the signal distributions are broad and overtake the SM background in the tails of the distribution ($m_{jj}$ values greater than 3 TeV). Figure~\ref{efficiency} shows the cumulative selection efficiency after each additional criteria outlined in Table \ref{fullcuts}. The signal acceptance is 0.8-1$\%$ depending on $m_{\chi}$, while the $W/Z$+jets backgrounds are reduced by approximately 6-7 orders of magnitude. Although we propose to determine final discovery potential with a shape based analysis (described later) using the full $m_{jj}$ spectrum, to illustrate where the bulk of the sensitivity lies, Figure \ref{mjj_optimization} shows the normalized signal significance $z/z_{max}$ as a function of the $m_{jj}$ cut value, assuming an integrated luminosity of 100 $fb^{-1}$.  As evidenced in Figure \ref{mjj_optimization}, the contribution to the total signal significance dominates for $m_{jj} > 3200$ GeV.

\section{Results} \label{results}
\begin{figure}
    \centering
    \includegraphics[width=.5\textwidth, height=.35\textheight]{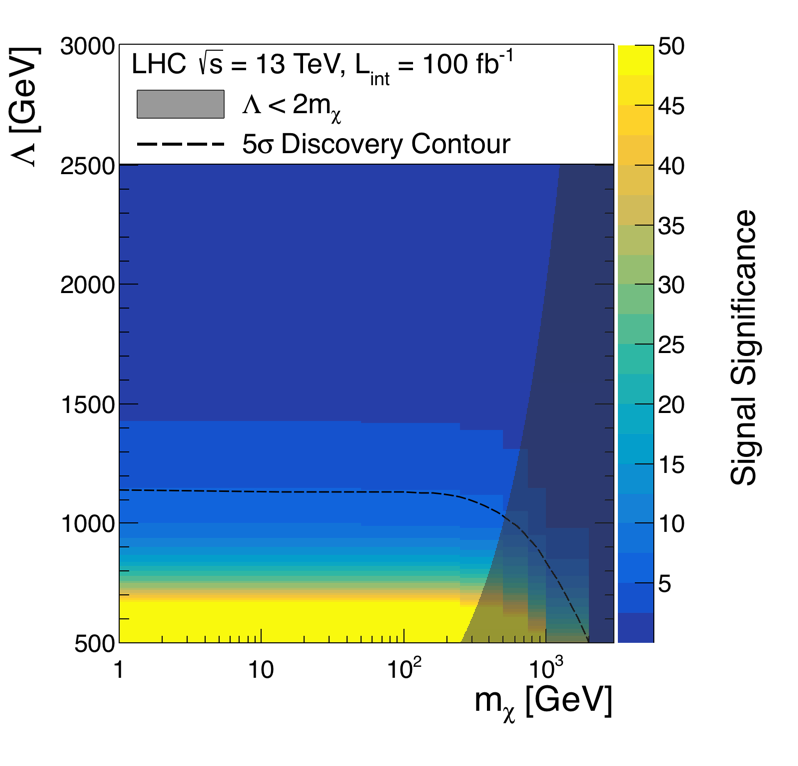}
    \caption{Expected signal significance as a function of the cutoff scale $\Lambda$ and the ADM mass $m_{\chi}$.  The signal significance was calculated by performing a profile binned likelihood of the $m_{jj}$ distribution using the systematic uncertainty as a nuisance parameter, assuming a luminosity of 100 $fb^{-1}$.  The $5\sigma$ discovery potential region is enclosed by the black dashed line, while the shaded grey area is the kinematically forbidden region $\Lambda < 2m_{\chi}$.}
    \label{moneyplot1}
\end{figure}
\begin{figure}
    \centering
    \includegraphics[width=.5\textwidth, height=.35\textheight]{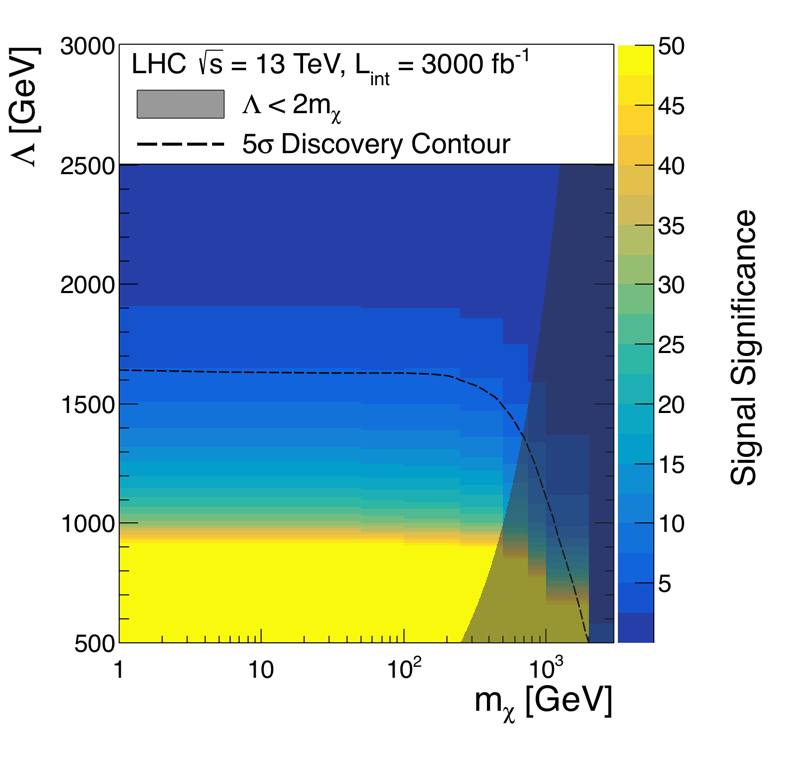}
    \caption{Expected signal significance as a function of the cutoff scale $\Lambda$ and the ADM mass $m_{\chi}$.  The signal significance was calculated by performing a profile binned likelihood of the $m_{jj}$ distribution using the systematic uncertainty as a nuisance parameter, assuming a luminosity of 3000 $fb^{-1}$.  The $5\sigma$ discovery potential region is enclosed by the black dashed line, while the shaded grey area is the kinematically forbidden region $\Lambda < 2m_{\chi}$.}
    \label{moneyplot2}
\end{figure}
The definitions for $S$, $B$, and signal significance $z$ described in the previous section are used only for the purpose of optimizing the selections. Instead of a cut and count approach, the full $m_{jj}$ distribution is used to perform a shape based determination of the discovery potential with the test statistic defined from a profile binned likelihood approach via the RooFit toolkit \cite{ROOTFit}.  The expected bin-by-bin yields of the $m_{T}$ distribution in Figure~\ref{dijetmass}, obtained using events satisfying the selections in Table I, are used as input to the profile binned likelihood calculation. These bin-by-bin yields are allowed to vary up and down based on their Poisson uncertainty and systematic uncertainty, where variations due to systematic uncertainty are incorporated using nuisance parameters. To incorporate realistic systematic uncertainties, we use the VBF SUSY searches at CMS as a guideline~\cite{CMSVBFDM, VBF2}. For the case of the SM backgrounds, the dominant systematic uncertainty in Ref.~\cite{CMSVBFDM, VBF2} is from the data-driven measurement of the VBF selection efficiency (24\%). The dominant source of systematic uncertainty in the expected signal yield comes from the experimental difficulties involved with reconstructing, identifying, and calibrating forward jets (20\%). Less significant contributions to the systematic uncertainties arise from the efficiencies for electron, muon, and $\tau_{h}$ identification (2-5\% depending on the lepton), which contribute due to the lepton vetos. Additionally, although we note a study of the appropriate trigger for the proposed VBF ADM search under the HL-LHC conditions is outside the scope of this work, we assume a 3\% systematic uncertainty on trigger efficiency following Ref.~\cite{CMSVBFDM, VBF2}. Finally, for the uncertainty due to the choice of parton distribution function (PDF) used to simulate the signal and background samples, we follow the PDF4LHC recommendations~\cite{Butterworth:2015oua}, resulting in a 5-15\% uncertainty depending on the process.  With the definition of the systematic uncertainties and determination of the fit variable, the final signal significance is calculated as the value of $z$ such that the integral of a Gaussian between $z$ and $\infty$ is equal to the probability that the background only test statistic value is comparable to that obtained with a signal plus background hypothesis.
\\\indent 
Based on the current data available at CMS/ATLAS and the projections for the HL-LHC, the  signal significance has been calculated for a range of luminosity values between 100 and 3000 $fb^{-1}$.  For each luminosity, we calculate the significance for various cutoff scales and ADM masses. The $m_{\chi}$ values range from 1 to 2000 GeV, while $\Lambda$ varies between 500 and 3000 GeV.  Figures \ref{moneyplot1} and \ref{moneyplot2} show the signal significance (on the z-axis) as a function of $m_{\chi}$ and $\Lambda$ on the $xy$-plane.  The region with   $\Lambda < 2m_{\chi}$, where the effective field theory breaks down, is shaded in gray.  
The black dashed line indicates the $5\sigma$ discovery contour (i.e.~$\{ \Lambda,m_{\chi} \}$ points below this line result in a signal significance $\ge 5\sigma$).
There is $5\sigma$ discovery potential for $m_{\chi}$ up to 600 GeV, assuming $\Lambda = 1$ TeV and an integrated luminosity of 100 $fb^{-1}$.  The discovery range for $m_{\chi}$, assuming $\Lambda = 1$ TeV  increases to $m_{\chi} < 1200$ GeV for an integrated luminosity of 3000 $fb^{-1}$.  It is important to note Figures~\ref{moneyplot1} and~\ref{moneyplot2} show that while the proposed VBF ADM search can probe TeV scale ADM masses, it may also achieve discovery potential for ``light" mass scenarios. For a light mass of $m_{\chi} = 1$ GeV, the proposed search can provide signal significances greater than $5\sigma$ for cutoff scales up to 1600 GeV, assuming an integrated luminosity of 3000 $fb^{-1}$.
\section{Discussion} \label{discussion}

In this work, we have explored the possibility of using VBF processes as probes to discover DM that couples to the SM through higher electromagnetic moments. Remaining agnostic about the UV completion, we consider the anapole DM operator within an effective field theory as a benchmark scenario. In this context, we denote $\chi$ as the ADM particle and study the discovery reach as a function of the free parameters $\Lambda$ and $m_{\chi}$, the cutoff scale and ADM mass, respectively. This EFT approach allows us to study a broad range of ADM masses, including very light DM scenarios (below 10 GeV and down to GeV or MeV scale). 

We find that $\gamma\gamma$ fusion is an important VBF ADM production mechanism, with cross sections that dominate over those of the more traditional mono-$X$ processes for all relevant values of $\Lambda$ and $m_{\chi}$. A particularly interesting feature resulting from $\gamma\gamma$ fusion events within the ADM EFT is that it leads to a VBF topology with significantly more forward jets and a larger dijet pseudorapidity gap compared to VBF DM production in other models such as SUSY, where t-channel WW/ZZ/WZ fusion diagrams dominate. This distinguishing feature of $\gamma\gamma$ fusion provides a clean mechanism to experimentally distinguish ADM from other DM scenarios should there be evidence for discovery with the LHC data. We have shown that the stringent requirements of large $E_{T}^{miss}$, two high $p_{T}$ very forward jets with unusually large separation in pseudorapidity, and TeV scale dijet mass is effective in reducing contributions from QCD multijet, $Z(\to{\nu\nu})$ + jets, $W(\to{l\nu})$ + jets, and other SM backgrounds.  

Assuming proton-proton collisions at $\sqrt{s} = 13$ TeV at the HL-LHC, the proposed VBF $\chi\chi{j}{j}$ search is expected to achieve a discovery reach with signal significance of at least 5$\sigma$ for ADM masses up to 1.2 (0.6) TeV and $\Lambda$ cutoff scales up to 1.65 TeV. For an example comparison with previously proposed ADM searches, the mono-$Z$ study of Ref.~\cite{Alves:2017uls} showed an expected discovery reach of $\Lambda \lessapprox 700$ GeV for $m_{\chi} = 100$ GeV and a similar range of systematic uncertainty. The proposed VBF ADM search in this Letter is expected to be the most important mode for discovery, far exceeding the projected sensitivity achievable by the mono-$X$ analyses.
\section{Acknowledgements}

We thank the constant and enduring financial support received for this project from the faculty of science at Universidad de los Andes (Bogot\'a, Colombia), the administrative department of science, technology and innovation of Colombia (COLCIENCIAS), the Physics \& Astronomy department at Vanderbilt University and the US National Science Foundation. This work is supported in part by NSF Award PHY-1806612 and a Vanderbilt Discovery Grant. KS is supported by the U.S. Department of Energy grant desc0009956.

\end{document}